\documentclass[a4paper,11pt]{article}

\usepackage{latexsym}
\usepackage{amsfonts} 
\usepackage[mathscr]{euscript}
\usepackage{amsmath,mathrsfs,bm,amssymb,color}

\title{\textsf{ Monotonicity of the polaron energy II:\\ General theory of
operator monotonicity
}}
\date{\empty}

\author{
Tadahiro Miyao\\ 
 {\it Department of Mathematics,}
{\it Hokkaido University,}\\
{\it Sapporo 060-0810, Japan}\\
 {\tt miyao@math.sci.hokudai.ac.jp}
}

\newcommand{\one}{{\mathchoice {\rm 1\mskip-4mu l} {\rm 1\mskip-4mu l}
{\rm 1\mskip-4.5mu l} {\rm 1\mskip-5mu l}}}
\newcommand{\h}{\mathfrak{h}}

\newcommand{\ex}{\mathrm{e}}
\newcommand{\D}{\mathrm{dom}}

\newcommand{\im}{\mathrm{i}}
\newcommand{\Fock}{\mathfrak{F}}
\newcommand{\Ffin}{\mathfrak{F}_{\mathrm{fin}}}

\newcommand{\dG}{\mathrm{d}\Gamma}

\newcommand{\ran}{\mathrm{ran}}

\newcommand{\la}{\langle}
\newcommand{\ra}{\rangle}

\newcommand{\BbbR}{\mathbb{R}}
\newcommand{\BbbN}{\mathbb{N}}

\newcommand{\BbbC}{\mathbb{C}}
\newcommand{\vepsilon}{\varepsilon}
\newcommand{\vphi}{\varphi}

\newcommand{\Pf}{P_{\mathrm{f}}}
\newcommand{\Nf}{N_{\mathrm{f}}}

\newcommand{\Cone}{\mathfrak{p}}

\newcommand{\dm}{\mathrm{d}}

\newcommand{\no}{\nonumber \\}

\def\Sumoplus{\sideset{}{^{\oplus}_{n\ge 0}}\sum}

\begin{document}

\newtheorem{define}{Definition}[section]
\newtheorem{Thm}[define]{Theorem}
\newtheorem{Prop}[define]{Proposition}
\newtheorem{lemm}[define]{Lemma}
\newtheorem{rem}[define]{Remark}
\newtheorem{assum}{Condition}
\newtheorem{example}{Example}
\newtheorem{coro}[define]{Corollary}

\maketitle

\begin{abstract}
We construct a general theory of operator monotonicity and apply it to
 the Fr\"ohlich polaron hamiltonian. This general theory provides a
 consistent viewpoint of the Fr\"ohlich model.
\end{abstract} 

\section{Introduction}
This paper is a sequel to  \cite{Miyao3}.
To explain our motivation of this work, let us recall the result in \cite{Miyao3} first.
In the previous work  \cite{Miyao3}, we studied the Fr\"ohlich hamiltonian  defined by 
\begin{align}
H_{\Lambda}= 
-\frac{1}{2}\Delta_x-\sqrt{\alpha}\lambda_0 \int_{|k|\le \Lambda}\dm k\, 
\frac{1}{|k|}
[\ex^{\im k\cdot x} a(k)+\ex^{-\im k\cdot x}a(k)^*]+\Nf.
\label{Polaron}
\end{align} 
Here $\Lambda$ is the ultraviolet cutoff. (The complete definiton of
$H_{\Lambda}$ will be recalled in the subsequent section.)
One of the  main results in \cite{Miyao3} is  stated as follows.

\begin{Thm}\label{ResultinI}
{\rm \cite{Miyao3}}
Let $E_{\Lambda}=\inf \mathrm{spec}(H_{\Lambda})$.
Then $E_{\Lambda}$ is strictly decreasing in $\Lambda$.
\end{Thm} 
In the proof of Theorem \ref{ResultinI}, a new operator monotonicity played important roles.
To illustrate  the meaning of our operator monotonicity, we
introduce some terms. 
 Let  $\Cone$ be a convex cone in the Hilbert space $\h$.
$\Cone$ is called to be {\it self-dual} if it satisfies
\begin{align}
\Cone=\{x\in \h\, |\, \la x, y\ra\ge 0 \, \forall y\in \Cone\}.
\end{align} 
Let $A, B$ be linear operators in $\h$. For simplicity, suppose both
are bounded. If $A$ and $B$ satisfy $(A-B)\Cone \subseteq \Cone$,
 we denote this as $A\unrhd B$ w.r.t. $\Cone$. 
It is easily  checked that the binary relation ``$\unrhd$'' is a partial 
order. Thus we can regard the relation $A\unrhd B$ w.r.t. $\Cone$ as an 
operator inequality. 
The operator monotonicity, our central subject in this paper, is expressed
as follows. Let $\{A_{s}\}_{s\ge 0}$ be a family of operators in $\h$.
We say $A_s$ is {\it monotonically decreasing} if 
\begin{align}
s>s'\ \ \Longrightarrow\ \ A_{s'}\unrhd A_s\ \ \mbox{w.r.t. $\Cone$}.
\end{align} 
To prove Theorem \ref{ResultinI}, we effectively applied this notion in
\cite{Miyao3}, namely, we selected a proper self-dual cone so that 
a family of hamiltonians $\{H_{\Lambda}\}_{\Lambda \ge 0}$  becomes monotonically
decreasing under   the choice.
(It will be seen that  we  can  extend  the notion of operator monotonicity to unbounded
operators, see Section \ref{SecMono}.)
 Combining this 
monotonicity and a general theorem established in \cite{Miyao3}, one
obtained the assertion in Theorem \ref{ResultinI}.
In this paper, we search for further possibilities of the operator
monotonicity. We construct a theory of the singular perturbation by the
operator monotonicity, and apply it to the Fr\"ohlich polaron hamiltonian.
The results in the part I\cite{Miyao3} are included in this theory. Our theory in
this paper provides a consistent viewpoint of the polaron as well.

Next we briefly background mathematical difficulties studying the
Fr\"ohlich polaron model and then state the results concerned these
 difficulties.
When we tackle the problem of investigating the mathematical aspects 
of the polaron model,  we unavoidably  encouter the question how we define 
 the hamiltonian. 
In various physical literatures,  one presupposes that  the hamiltonian without the ultraviolet
cutoff $H_{\Lambda=\infty}$   is given.
However it is not obvious whether the hamiltonian $H_{\Lambda=\infty}$
 is definable or not.
To clarify the point, put $\Lambda=\infty$ in the electron-phonon
interaction term in (\ref{Polaron}). Then since $1/|k|$ is not in 
$L^2(\BbbR^3)$, the interaction term can not be defined as a linear
operator in the Hilbert space.
(Remark the symbols 
$\int_{\BbbR^3}\dm k\, f(k) a(k)$ and $\int_{\BbbR^3} \dm k f(k) a(k)^*$
are well-defined linear operators only if $f\in L^2(\BbbR^3)$.
)
If one looks at only interaction term,  the interaction term 
does not have mathematical meaning at $\Lambda=\infty$, however
if our vison   broaden  and  we study the whole hamiltonian including 
not only the electron-phonon interaction term but also the electron and
phonon kinetic  energy terms,  the limit $\lim_{\Lambda\to \infty}
H_{\Lambda}$ exists in a suitable sense.
This is the standard way how to define the Fr\"ohlich hamiltonian 
\cite{Eck, JFroehlich1,JFroehlich2,  GeLowen,LGross2,  Nelson, Pizzo, Sloan2}.
In this paper, we prove the existence of the limiting hamiltonian 
by applying the general theory of the operator monotonicity.
As far as we know, our proof is novel.
In addition,  this provides an important  example of our operator monotonicity.

Next problem is to study the ground state property of the hamiltonian
without the ultraviolet cutoff. 
(Here keep in mind that we are discussing the ground state of the
hamiltonian at a fixed total momentum now.)
Remark that existence of the ground state has been  already established 
\cite{JFroehlich1, GeLowen, Moller2, Spohn2}.
In this paper we argue  the uniqueness of the ground state.
 This is known to be rather difficult because the hamiltonian is 
defined through the limiting procedures. 
In \cite{Miyao}, the author proved the uniqueness of the ground
state. In the present work, we will give a  different proof by the 
general theory of the operator monotonicity.

Once again, let us  reconfirm the aim of this work.
Through analysis of the Fr\"ohlich hamiltonian, we build a general
theory of 
the operator monotonicity.
We expect the further validity of our operator inequalities is  revealed
through our attempt.

The organization of the paper is as follows.
In Sects. 2 and 3, we establish a general theory of our operator
monotonicity.
Terminologies from the self-dual cone analysis \cite{Miyao, Miyao2, Miyao3,
Miyao4, Miyao5} are introduced in this
section.
In Sect. 4, we summarize basic facts of the second quantization.
Sect. 5 is devoted to display applications of the general theory  in
Sects. 2 and 3 to the Fr\"ohlich polaron model. 
In Sects. 6-10, we give detailed proofs of the results  stated in Sect 5.

\begin{flushleft}
{\bf Acknowledgments}\\
I  would like to thank H. Spohn for useful discussions.
Financial support by KAKENHI (20554421) is gratefully acknowledged.
\end{flushleft}

\section{A general theory of operator monotonicity}\label{SecMono}
\setcounter{equation}{0}

In this section, we will provide an abstract theory of operator
monotonicity.
As we will see in the later sections,  this abstract theory has an
important application to the condensed matter physics.

\subsection{Definitions}\label{Def}
To express our ideas, results and proofs, we need to introduce some
technical terms from an  earlier work \cite{Miyao3}.
The terms defined here serve as a language  in this paper.

Let $\h$ be a complex  Hilbert space  and $\Cone$ be a convex cone in
$\h$.
Then $\Cone $ is called to be {\it self-dual} if 
\begin{align}
\Cone=\{x\in \h\, |\, \la x, y\ra \ge 0\  
\forall y\in \Cone
\}.
\end{align} 
Henceforth $\Cone$ always  denotes  the self-dual cone in $\h$.
The following properties of $\Cone$ are well-known \cite{Bos, Haagerup}:
\begin{Prop}\label{BasisSAC}One has the following.
\begin{itemize}
\item[{\rm (i)}] $\Cone\cap (-\Cone)=\{0\}$.
\item[{\rm (ii)}] There exists a unique involution $j$ in $\h$ such that
                 $jx=x$ for all $x\in \Cone$.
\item[{\rm (iii)}] Each element $x\in \h$ with $jx=x$ has a unique
                 decomposition $x=x_+-x_-$,  where $x_+,x_-\in\Cone$ and
                 $\la x_+, x_-\ra=0$.
\item[{\rm (iv)}] $\h$ is linearly spanned by $\Cone$.
\end{itemize} 
\end{Prop} 
\begin{rem}\label{LinearSpan}
{\rm
We clarify the precise meaning of (iv) in Proposition \ref{BasisSAC}.
Each $x\in \h$ can be expressed as $x=\Re x+\im \Im x$, where 
$\Re x=\frac{1}{2}(\one +j)x$ and $\Im x=\frac{1}{2\im}(\one-j)x$.
Clearly $j\Re x=\Re x $ and $j\Im x=\Im x$. Thus, by (iii), we have a
 unique decomposition $\Re x=(\Re x)_+-(\Re x)_-$ with $\la (\Re x)_+,
 (\Re x)_-\ra=0$. Similar property holds for $\Im x$. Now one has 
\begin{align}
x=(\Re x)_+-(\Re x)_-+\im \{(\Im x)_+- (\Im x)_-\}.
\end{align} 
Of course, $(\Re x)_{\pm}, (\Im x)_{\pm}\in \Cone$.
This is the meaning of (iv). $\diamondsuit$
}
\end{rem}

If $x-y \in \Cone$, then we will write $x\ge y$ (or $y\le x$)
 w.r.t. $\Cone$.
Let $A$ and $B$ be densely defined linear operators on $\h$.
If $Ax \ge Bx$ w.r.t. $\Cone $ for all $x\in \D(A)\cap \D(B)\cap \Cone$,
then we will write $A\unrhd B$ (or $B\unlhd A$) w.r.t. $\Cone$.
Especially if $A$ satisfies $0\unlhd A$  w.r.t. $\Cone$, then we say
that $A$ {\it preserves positivity with respect to }$\Cone$.
The symbol ``$\unrhd$'' was first introduced by Miura
 \cite{Miura}.

An element $x$ in $\Cone$ is called to be {\it strictly positive } if
$\la x, y \ra\ge 0$ for all $y\in \Cone\backslash \{0\}$.
We will  write this as $x>0$ w.r.t. $\Cone$. Of course, an inequality  $x>y$
w.r.t. $\Cone $ means $x-y$ is strictly positive w.r.t. $\Cone$.
If bounded operators $A$ and $B$ satsify $Ax > Bx$ w.r.t. $\Cone$ for
all $x\in \Cone\backslash\{0\}$, then we will express this as $A\rhd B$
(or $B\lhd A$) w.r.t. $\Cone$. Clearly if $A\rhd B$ w.r.t. $\Cone$, then
$A\unrhd B$ w.r.t. $\Cone$. We say that $A$ {\it improves positivity
w.r.t. $\Cone$} if $A\rhd 0$ w.r.t. $\Cone$.

\subsection{Monotonically decreasing self-adjoint operators}

Let $\Cone$ be a self-dual cone in the Hilbert space $\h$.
Let $\{H_{\lambda}\}_{\lambda\ge 0}$  be a family  of self-adjoint operators on
$\h$. In this section we always assume the following.

\begin{itemize}

\item[{ \bf(A. 1)}] There exists a constant $M>-\infty$, independent of
	     $\lambda$,
such that $H_{\lambda} \ge M$ for all $\lambda \ge 0$.

\item[{\bf (A. 2)}] For all $\lambda \ge 0$ and $s \ge 0$,
$\displaystyle 
\ex^{-s H_{\lambda} } \unrhd 0
$ w.r.t. $\Cone$. 
\item[{\bf (A. 3)}] Each $H_{\lambda}$ has the  common domain $\mathscr{D}$, i.e., 
$\D(H_{\lambda})=\mathscr{D}$ for all $\lambda \ge 0$.
\end{itemize}

Under these assumptions, we can show the existence of the limiting
hamiltonian as follows.

\begin{Thm}\label{Existence}
We assume (A. 1), (A. 2) and (A. 3). Suppose that $H_{\lambda}$ is 
 monotonically  decreasing  in $\lambda$ in the sense that 
\begin{align}
\lambda_1 \le \lambda_2 \Longrightarrow H_{\lambda_1}\unrhd H_{\lambda_2} \  \ \ \mbox{w.r.t. $\Cone$.}\label{Monotonicity}
\end{align}
Then, there exists a unique self-adjoint operator $H$, bounded from
 below
by $M$, with following properties:
\begin{itemize}
\item[{\rm (i)}] $H_{\lambda}$ converges to $H$ in strong resolvent sense as $\lambda\to
	   \infty$,
\item[{\rm (ii)}] For all $\lambda \ge 0$ and $s\ge 0$, $\ex^{-s H}\unrhd \ex^{-s H_{\lambda}}
$ w.r.t. $\Cone$. In particular, $\ex^{-s H}\unrhd 0$
w.r.t. $\Cone$ for all $s\ge 0$. 
\end{itemize} 
\end{Thm}

In the above theorem, we proved   the limiting hamiltonian denoted by
$H$
 satisfies $\ex^{-s H}\unrhd 0$ w.r.t. $\Cone$ for all $s\ge 0$.
Our next problem is to show $\ex^{-s H} \rhd 0$ w.r.t. $\Cone$ for all
$s>0$.
This problem is important  because if we can solve it, then by the
Perron-Frobenius-Faris theorem(Theorem \ref{Faris}), the ground state of $H$ (if it exsits)
 is automatically unique and strictly positive w.r.t. $\Cone$.
Below we will give two answers about this problem.

\begin{Thm}\label{PositivityImp}
In addition to the assumptions in Theorem \ref{Existence},
assume that, there exists a $\lambda \ge 0$ such that 
\begin{align}
\ex^{-s H_{\lambda}} \rhd 0\label{PIn}
\end{align} 
w.r.t. $\Cone$ for all $s>0$. Let $H$ be  the self-adjoint operator 
obtained in Theorem \ref{Existence}.
Then one has  
\begin{align}
\ex^{-s H} \rhd 0\label{PI}
\end{align} 
 w.r.t. $\Cone$ for all $s>0$. Consequently, if $\inf \mathrm{spec}(H)$
 is an eigenvalue, then it is unique and the corresponding eigenvector 
 can be chosen as  strictly positive w.r.t. $\Cone$.
\end{Thm} 

Theorem \ref{PositivityImp} was  applied  to show the uniqueness of the ground
state for the polaron and bipolaron Hamiltonians without cutoffs  in
\cite{MS, Miyao}.
Instead,  
 the following theorem is convenient for  concrete applications in this
 paper.

\begin{Thm}\label{Ergodic}
In addition to the assumptions in Theorem \ref{Existence},
assume that, for some $\mu\ge 0$, a family  of self-adjoint operators
$\{(H_{\lambda}+\mu)^{-1}\}_{\lambda\ge 0}$ is ergodic in the sense that, for any $x, y\in
 \Cone\backslash
\{0\}$, there exists a $\lambda \ge 0$ so that $\la x, (H_{\lambda}+\mu)^{-1}y\ra>0$. 
 Let $H$ be  the self-adjoint operator  in Theorem \ref{Existence}.
Then one has  
\begin{align}
\ex^{-s H} \rhd 0\label{PI}
\end{align} 
 w.r.t. $\Cone$ for all $s>0$. Consequently, if $\inf \mathrm{spec}(H)$
 is an eigenvalue, then it is unique and the corresponding eigenvector 
 can be chosen as  strictly positive w.r.t. $\Cone$.
\end{Thm} 

The operator  monotonicity (\ref{Monotonicity}) further gives  useful information
concerning spectrum  of the Hamiltonians.
Indeed, in our previous work \cite{Miyao3}, the following 
theorem was proven.

\begin{Thm}{\rm \cite{Miyao3}}\label{EnergyMono}
Assume (A. 1), (A. 2), (A. 3),  (\ref{Monotonicity}).
Set $E_{\lambda}=\inf \mathrm{spec}(H_{\lambda})$ and $E=\inf \mathrm{spec}(H)$.
Then $E_\lambda$ is monotonically decreasing in $\lambda$ and
 $\displaystyle \lim_{\lambda \to \infty} E_{\lambda}=E$.
\end{Thm} 

\section{Proofs of Theorems \ref{Existence},  \ref{PositivityImp},
 \ref{Ergodic}}

\subsection{Proof of Theorem \ref{Existence}}

{\bf STEP 1.} Let $x, y\in \Cone$. Define a function $F_{x,y}(\lambda)$
 by $F_{x, y}(\lambda)=\la x, \ex^{-H_{\lambda}}y\ra$.
 Then, by (\ref{Monotonicity}), the functiom  
 $F_{x, y}(\lambda)$  is   monotonically 
increasing in $\lambda$  and, by the assumption  {\bf (A. 1)}, it is bounded. Hence
 $\displaystyle \lim_{\lambda\to \infty}F_{x, y}(\lambda)$
exists. Then, since $\h$ is linearly spanned by $\Cone$(Remark \ref{LinearSpan}), the limit
$\displaystyle 
\lim_{\lambda\to
\infty} \la x, \ex^{-s H_{\lambda}}y\ra$ exists for all $x, y\in \h$. 
Now one can define a sesquilinear  form $B_s(\cdot, \cdot)$ by 
\begin{align}
B_s(x, y)=\lim_{\lambda\to \infty} \la x, \ex^{-s H_{\lambda}}y\ra.
\end{align} 
By the assumption {\bf (A. 1)}, one sees
$
|B_s(x, y)|\le \|x\| \|y\|\,  \ex^{-s M}.
$
Thus, by the Riesz's representation theorem, there exists a unique positive
operator $T_s$ such that 
\begin{align}
B_s(x, y)=\la x, T_s y\ra.
\end{align} 
Then, by definition, one has 
\begin{align}
\|T_s\| \le \ex^{-s M}, \ \ \ 
T_s= \mbox{w-}\lim_{\lambda \to \infty}\ex^{-s H_{\lambda}}, \label{WeakLim}
\end{align} 
where $\mbox{w-}\lim$ means the weak limit. Moreover, by the
monotonicity (\ref{Monotonicity}), one has 
\begin{align}
\ex^{-s H_{\lambda}} \unlhd T_s \label{UpperBd}
\end{align} 
for all $\lambda \ge 0$. [Proof:
By Proposition  \ref{ABIneqEq}, one has $\ex^{-s H_{\lambda_2}} \unlhd \ex^{-s H_{\lambda_2}}$
 whenever $\lambda_2\ge\lambda_1$.  Then taking $\lambda_2\to \infty$, one concludes
(\ref{UpperBd}). Remark that to apply Proposition \ref{ABIneqEq}, we used
the assumptions {\bf (A. 2)} and {\bf (A. 3)}.
]
\medskip\\
{\bf STEP 2.} In this step we will show
\begin{align}
\mbox{s-}\lim_{\lambda\to \infty}\ex^{-s H_{\lambda}}=T_s, \label{StrongLim}
\end{align} 
where $\mbox{s-}\lim$ means the strong limit. 

Since $\h$ is linearly spanned by $\Cone$(Remark \ref{LinearSpan}), it sufficies to show that 
\begin{align}
\mbox{s-}\lim_{\lambda\to \infty}\ex^{-s H_{\lambda }}x=T_s x\label{ConeOnly}
\end{align}
for all $x\in \Cone$.
To this end, observe that 
\begin{align}
\|
(T_s-\ex^{-s H_{\lambda}})x
\|^2
=\big\la x, \big(T_s^2
-\ex^{-s H_{\lambda}}T_s-T_s \ex^{-sH_{\lambda}}+\ex^{-sH_{\lambda}} \ex^{-sH_{\lambda}}
\big)x\big\ra.\label{StrongConv}
\end{align} 
Note that, by (\ref{UpperBd}), one has 
$\ex^{-sH_\lambda}\ex^{-s H_{\lambda}}\unlhd T_s \ex^{-s H_{\lambda}}$ which implies 
$\la x,  \ex^{-s H_{\lambda}}\ex^{-s H_{\lambda}}x\ra \le \la x, T_s \ex^{-s H_{\lambda}}x\ra$.
Hence
\begin{align}
\mbox{RHS of (\ref{StrongConv})}
&\le \big\la x, \big(
T_s^2-\ex^{-s H_{\lambda}} T_s-T_s \ex^{-s H_{\lambda}}+T_s \ex^{-s H_{\lambda}}
\big)x\big\ra\no  
&=\la x, (T_s-\ex^{-s H_{\lambda}})T_s x\ra\no
&\to 0
\end{align} 
as $\lambda\to \infty $ by (\ref{WeakLim}). This proves (\ref{ConeOnly}).
\medskip\\
{\bf STEP 3.}
Since $\{\ex^{-s H_{\lambda}}\}_{s\ge 0}$ is a one-parameter semigroup,
$\{T_s\}_{s\ge 0}$ is also one-parameter semigroup  by STEP 2.
If one can show the strong continuity of $T_s$ in $s\ge 0$,  
 one sees that there exists a unique
self-adjoint operator $H$ such that $T_s=\ex^{-s H}$ by Propositon \ref{SemiG}. By our
construction,
 it is clear that 
\begin{align}
\ex^{-sH}= \mbox{s-}\lim_{\lambda\to \infty}\ex^{-s H_{\lambda}},\ \ \
\ex^{-s H}\unrhd \ex^{-s H_{\lambda}},\ \ \
H \ge M.
\end{align} 
This is the assertion in the theorem. 

Let  prove the strong continuity of $T_s$ in $s$.
Without loss of generality, we may assume $M>0$, i.e., $H_{\lambda}\ge
M>0$. (Indeed, if $M>0$, there is nothing to do. On the other hand, if $M\le
0$, then we simply study
$\tilde{H}_{\lambda}=H_{\lambda}-M+\vepsilon\ (\vepsilon>0)$ instead of
$H_{\lambda}$ itself. Obviously $\tilde{H}_{\lambda}\ge \vepsilon$ for
all $\lambda$. ) 
Hence $\one -T_s \ge 0$ holds for all $s \ge 0$.
Fix $\lambda \ge 0$ arbitrarily. One has, by (\ref{UpperBd}),
\begin{align}
\ex^{-s H_{\lambda}} \unlhd T_s
\end{align} 
w.r.t. $\Cone$. Thus $\one-\ex^{-s H_{\lambda}} \unrhd \one -T_s$ w.r.t. $\Cone$.
Hence, for all $x \in \Cone$, we have 
\begin{align}
0\le \la x, (\one-T_s) x\ra\le \la x, (\one -\ex^{-s H_{\lambda}})x\ra
\end{align} 
for all $s \ge 0$.
Taking $s\ \to +0$, 
\begin{align}
\lim_{s\to +0} \la x, (\one-T_s)x\ra=0\label{WeakLimP}
\end{align} 
holds for all $x\in \Cone$.
 Next observe that, for any $x\in
\Cone$,
\begin{align}
\|(\one-T_s) x\|^2=\la x, (\one-2T_s+T_{2s})x\ra \to 0
\end{align} 
as $s\to +0$ by (\ref{WeakLimP}).
Then since $\h$ is linearly spanned by $\Cone$(Remark \ref{LinearSpan}), one sees $\|(T_s-\one)x\|\to
0$ as $s\to +0$ for all $x\in \h$.
$\Box$

\subsection{Proof of Theorem \ref{PositivityImp}}

The proof of this theorem was already given in \cite{Miyao}.
Here we will repeat it for reader's convenience.

By  Theorem \ref{Existence} (ii),
\begin{align}
\ex^{-s H}\unrhd \ex^{-s H_{\lambda}}
\end{align} 
w.r.t. $\Cone$. Since $\ex^{-s H_{\lambda}}\rhd 0$,  one arrives
at 
\begin{align}
\ex^{-s H}\unrhd \ex^{-s H_{\lambda}} \rhd 0
\end{align} 
 w.r.t. $\Cone$ for all $s>0$. Remainder assertions comes from
 Theorem \ref{Faris}. $\Box$

\subsection{Proof of Theorem \ref{Ergodic}}
By (\ref{Monotonicity}) and Proposition \ref{ABIneqEq}, one has 
\begin{align}
(H+\mu)^{-1} \unrhd (H_{\lambda}+\mu)^{-1}\label{ResolventInq}
\end{align}
w.r.t. $\Cone$ for all $\lambda \ge 0$. Choose $x, y \in \Cone \backslash
\{0\}$
arbitrarily. Then, by the ergodicity of
$\{(H_{\lambda}+\mu)^{-1}\}_{\lambda\ge 0}$, there is
a $\lambda \ge 0$ such that $\la x, (H_{\lambda}+\mu)^{-1}y\ra>0$. Combining this
with (\ref{ResolventInq}), we have 
\begin{align}
\la x, (H+\mu)^{-1}y\ra \ge \la x, (H_{\lambda}+\mu)^{-1}y\ra >0.
\end{align} 
This means $(H+\mu)^{-1}\rhd 0$ w.r.t. $\Cone$. Now we apply Theorem
\ref{Faris}
and  conclude the assertion. $\Box$

\section{Second quantization}\label{2ndQuant}
\setcounter{equation}{0}

Before we study the Fr\"ohlich polaron, we briefly summarize necessary
results of the second quantization. Many of these have already been
stated in the previous work \cite{Miyao3}.

\subsection{Basic definitions}
The bosonic  Fock space over $\h$ is defined by 
\begin{align}
\Fock(\h)=\Sumoplus 
\Fock^{(n)}(\h),
\ \ \ \Fock^{(n)}(\h)=\h^{\otimes_{\mathrm{s}}n},
\end{align} 
where $\h^{\otimes_{\mathrm{s}}n}$  is the $n$-fold
symmetric tensor product of $\h$ with convention 
$\h^{\otimes_{\mathrm{s}}0}=\BbbC$. 
$\Fock^{(n)}(\h)$ is called the {\it $n$-boson subspace}.

We denote by $a(f)\, (f\in \h)$ the annihilation operator on
$\Fock(\h)$, its adjoint $a(f)^*$, called the creation operator, is defined by
\begin{align}
a(f)^*\vphi=\sideset{}{^{\oplus}_{n\ge 1}}\sum
 \sqrt{n}S_n (f\otimes \vphi^{(n-1)})\label{DefCrea}
\end{align} 
for $\vphi=\sum_{n\ge 0}^{\oplus} \vphi^{(n)}\in \D(a(f)^*)$, where
$S_n$ is the symmetrizer  on $\Fock^{(n)}(\h)=\h^{\otimes_{\mathrm{s}}n }$.
The annihilation- and creation operators satisfy
the cannonical commutation relations (CCRs)
\begin{align}
[a(f), a(g)^*]=\la f, g\ra, \ \ [a(f), a(g)]=0=[a(f)^*, a(g)^*]
\end{align} 
on a suitable dense subspace in $\Fock(\h)$.

Let $C$ be a contraction operator on $\h$, that is ,
$\|C\|\le 1$. Then we define a contraction operator $\Gamma(C)$ on
$\Fock(\h)$
by 
\begin{align}
\Gamma(C)=\Sumoplus C^{\otimes n}
\end{align} 
with $C^{\otimes 0}=\one$, the identity operator. 
For a self-adjoint operator $A$ on $\h$, let us introduce 
\begin{align}
\dG(A)=0\oplus \sideset{}{^{\oplus}_{n\ge 1}}\sum
 \sideset{}{_{n\ge k\ge 1}}\sum
\one^{\otimes (k-1)}\otimes A \otimes \one^{\otimes (n-k)}  
\end{align} 
acting in   $\Fock(\h)$. Then  $\dG(A)$ is essentially self-adjoint. We denote its
closure by the same symbol. A typical example is the bosonic number operator $\Nf=\dG(\one)$.
We remark the following relation between
$\Gamma(\cdot)$ and $\dG(\cdot)$:
\begin{align}
\Gamma(\ex^{\im t A})= \ex^{\im t \dG(A)}.
\end{align} 
In particular if $A$ is positive, then one has 
\begin{align}
\Gamma(\ex^{-tA})=\ex^{-t \dG(A)}.
\end{align} 

\begin{Prop}
Let $A$ be a positive self-adjoint operator. Then we have the following operator inequalities:
\begin{align}
a(f)^*a(f)&\le \|A^{-1/2}f\|^2 (\dG(A)+\one),\label{CreAnnInq}\\
\dG(A)+a(f)+a(f)^*&\ge -\|A^{-1/2}f\|^2. \label{VHove}
\end{align} 
\end{Prop} 
\subsection{Fock space over $L^2$-space}
In this paper, the bosonic Fock space over $L^2(\BbbR^3_k)=L^2(\BbbR^3,
\dm k)$ will often appear and we simply denote as 
\begin{align}
\Fock=\Fock(L^2(\BbbR^3_k)).
\end{align} 
The $n$-boson subspace $\Fock^{(n)}=L^2(\BbbR^3_k)^{\otimes_{\mathrm{s}}n}$ is naturally
identified with $
L^2_{\mathrm{sym}}(\BbbR^{3n})=\big\{
\vphi\in L^2(\BbbR^{3n}_k)\, |\, \vphi(k_1,\dots,
k_n)=\vphi(k_{\sigma(1)}, \dots, k_{\sigma(n)})\ \mbox{a.e. }\forall
\sigma \in \mathfrak{S}_n 
\big\}
$, where $\mathfrak{S}_n$ is the permutation group on a set $\{1, 2,
\dots , n\}$.
Hence 
\begin{align}
\Fock=\BbbC \oplus \sideset{}{^{\oplus}_{n\ge 1}}\sum
 L^2_{\mathrm{sym}}(\BbbR^{3n}_k). \label{FockId}
\end{align}
The annihilation- and creation operators are symbolically expressed as 
\begin{align}
a(f)=\int_{\BbbR^3}\dm k\, \overline{f(k)}a(k),\ \
 a(f)^*=\int_{\BbbR^3}\dm k\, f(k) a(k)^*.
\end{align}
If $\omega$ is a multipilication operator by the function $\omega(k)$,
then $\dG(\omega)$ is formally written as 
\begin{align}
\dG(\omega)=\int_{\BbbR^3_k}\dm k\, \omega(k)a(k)^* a(k).
\end{align}  

\subsection{The Fr\"ohlich cone}
In order to discuss the inequalities introduced in \S \ref{SecMono},  
we have to determine a   self-dual cone in $\Fock$.
Here we will intoduce  a  natural self-dual cone in $\Fock$ which is 
 suitable for our analysis in later sections.

Under  the natural identification
$L^2(\BbbR^3_k)^{\otimes_{\mathrm{s}}n}=L^2_{\mathrm{sym}}(\BbbR^{3n}_k)$,
set 
\begin{align}
\Fock_+^{(n)}=\big\{
\vphi \in L^2_{\mathrm{sym}}(\BbbR^{3n})\, |\, \vphi(k_1,\dots, k_n)\ge
 0\ \mbox{a.e.}
\big\}\label{FockIdentify}
\end{align}
with $\Fock^{(0)}_+=\BbbR_+=\{r\in \BbbR\, |\, r\ge 0\}$. Then each 
$\Fock_+^{(n)}$ is a self-dual cone in
$\Fock^{(n)}=L^2_{\mathrm{sym}}(\BbbR^{3n}_k)$.

\begin{define}
{\rm
The {\it Fr\"ohlich cone  } is defined by
\begin{align}
\Fock_+=\Sumoplus \Fock^{(n)}_+.
\end{align} 
Again $\Fock_+$ is a self-dual cone in $\Fock$. $\diamondsuit$
}
\end{define} 

We summarize  properties of operators in $\Fock$ below.
All propositions were proven in  \cite{Miyao3}.
\begin{Prop}
\label{PPFockI}
 Let $C$ be a contraction on $L^2(\BbbR^3_k)$.  Then if
$C\unrhd 0$ w.r.t. $L^2(\BbbR^3_k)_+$,   one has 
$\Gamma(C)\unrhd 0$ w.r.t. $\Fock_+$, where $L^2(\BbbR^3_k)=\{f\in
 L^2(\BbbR^3_k)\, |\, f(k)\ge 0\ a.e.\}$.
 Especially one has the following.
\begin{itemize}
\item[{\rm (i)}] For a self-adjoint operator $A$, if $\ex^{\im t A}\unrhd 0$
w.r.t. $L^2(\BbbR^3)_+$, then one has $\Gamma(\ex^{\im t A})\unrhd 0$
w.r.t. $\Fock_+$.
\item[{\rm (ii)}]  For a positive  self-adjoint operator $B$, if $\ex^{- t B}\unrhd 0$
w.r.t. $L^2(\BbbR^3)_+$, then one has $\Gamma(\ex^{-t B})\unrhd 0$
w.r.t. $\Fock_+$.
\end{itemize} 
\end{Prop}

\begin{Prop}\label{PPFockII}
If $f \ge 0$ w.r.t. $L^2(\BbbR^3_k)_+$, then $a(f)^*\unrhd 0$ and
 $a(f)\unrhd 0$ w.r.t. $\Fock_+$.
\end{Prop}

\begin{Prop}\label{ErgoFock}{\rm (Ergodicity)}
For each $f\in L^2(\BbbR^3_k)$, let $\phi(f)$ be a  linear
 operator
 defined by
\begin{align}
\phi(f)=a(f)+a(f)^*.
\end{align} 
If $f >0$ w.r.t. $L^2(\BbbR^3_k)_+$, that is, $f(k)>0$ a.e. $k$, then
$\phi(f)$ is ergodic in the sense  that, for any $x, y\in
 (\Fock_+\cap \Ffin) \backslash \{0\}$, there exists an $n\in \{0\}\cup\BbbN$ such
 that $\la x, \phi(f)^n y\ra >0$.
\end{Prop}

\subsection{Local properties}
Let $B_{\Lambda}$ be a ball of radius $\Lambda$ in $\BbbR^3_k$ and let
$\chi_{\Lambda}$ be a function on $\BbbR^3$ defined by
$\chi_{\Lambda}(k)=1$ if $k \in B_{\Lambda}$ and $\chi_{\Lambda}(k)=0$
otherwise. 
Then as a multiplication operator, $\chi_{\Lambda}$
is an orthogonal projection on $L^2(\BbbR^3_k)$ and
 $Q_{\Lambda}=\Gamma(\chi_{\Lambda})$ is also an orthogonal projection  on $\Fock$.
Now let us define the local Fock space by
\begin{align}
\Fock_{\Lambda}=Q_{\Lambda}\Fock.
\end{align} 
Clearly $\Fock=\Fock_{\infty}$. Since $\chi_{\Lambda}L^2(\BbbR^3_k)=L^2(B_{\Lambda})$, $\Fock_{\Lambda}$ can be
identified with $\Fock(L^2(B_{\Lambda}))$.

The following proposition was unstated in \cite{Miyao3}.
\begin{Prop}\label{QPP}
For each $\Lambda \ge 0$, put
 $Q_{\Lambda}^{\perp}=\one-Q_{\Lambda}$. Then one obtains the following.
\begin{itemize}
\item[{\rm (i)}] $Q_{\Lambda}\unrhd 0$ w.r.t. $\Fock_+$.
\item[{\rm (ii)}] $Q_{\Lambda}^{\perp} \unrhd 0$ w.r.t. $\Fock_+$.
\end{itemize} 
\end{Prop} 
{\it Proof.} (i) immediately follows from Proposition \ref{PPFockI}.

(ii) Under the identification (\ref{FockId}), we see
\begin{align}
(Q_{\Lambda}\vphi_n)(k_1, \dots, k_n)=\prod_{j=1}^n\chi_{\Lambda}(k_j)
 \vphi_n(k_1,\dots, k_n)
\end{align} 
for each $\vphi_n\in L^2_{\mathrm{sym}}(\BbbR^{3n}_k)$. Hence 
\begin{align}
(Q^{\perp}_{\Lambda}\vphi_n)(k_1,\dots, k_n)=\Big\{1-\prod_{j=1}^n
 \chi_{\Lambda}(k_j)\Big\} \vphi_n(k_1,\dots, k_n). \label{OrthQ}
\end{align} 
If $\vphi_n(k_1,\dots, k_n)\ge 0$ a.e., then the right hand side of
(\ref{OrthQ}) is positive for a.e. $k_1, \dots, k_n$  because
$1-\prod_{j=1}^n \chi_{\Lambda}(k_j)\ge 0$. This means
$Q^{\perp}_{\Lambda} \unrhd 0$ w.r.t. $\Fock_{+}$. $\Box$
\medskip\\

As to the annihilation- and creation operators, we remark the following
properties:
\begin{align}
a(f)Q_{\Lambda}&=a(\chi_{\Lambda}f)=\int_{|k|\le \Lambda}\dm k\,
 \overline{f(k)} a(k),\\
Q_{\Lambda} a(f)^*&=a(\chi_{\Lambda}f)^*=\int_{|k|\le \Lambda}\dm k\,
 f(k) a(k)^*,\\
\dG(\omega) Q_{\Lambda}&=\dG(\chi_{\Lambda}\omega)=\int_{|k|\le \Lambda}\dm k\, \omega(k)a(k)^*a(k).
\end{align}

Next let us introduce  a  natural self-dual cone in $\Fock_{\Lambda}$.
To this end, define
\begin{align}
\Fock_{\Lambda, +}^{(n)}=\big\{
\vphi\in L^2_{\mathrm{sym}}(B_{\Lambda}^{\times n})\, |\,
 \vphi(k_1,\dots k_n) \ge 0\ \  \mbox{a.e.}
\big\}
\end{align}
with $\Fock_{\Lambda, +}^{(0)}=\BbbR^+$.  Each $\Fock_{\Lambda,
+}^{(n)}$ is a self-dual cone in
$L^2(B_{\Lambda})^{\otimes_{\mathrm{s}}n}=L^2_{\mathrm{sym}}(B_{\Lambda}^{\times
n})$.
\begin{define}
{\rm
The {\it  local Fr\"ohlich cone } is defined by 
\begin{align}
\Fock_{\Lambda, +}=\Sumoplus \Fock_{\Lambda, +}^{(n)}.
\end{align} 
$\Fock_{\Lambda, +}$ is a self-dual cone in $\Fock_{\Lambda}$. $\diamondsuit$
}
\end{define} 

\begin{Prop}\label{LocalPropErgo}
Propositions \ref{PPFockI}, \ref{PPFockII} and \ref{ErgoFock} are still
 true even if one replaces $L^2(\BbbR^3_k)_+$ and $\Fock_+$ by
 $L^2(B_{\Lambda})_+$ and $\Fock_{\Lambda, +}$ respectively. 
\end{Prop} 
As to the proof of the above proposition, see \cite{Miyao3}.

\section{The  polaron: From a viewpoint of operator inequalities}

\setcounter{equation}{0}

\subsection{Definition of Hamiltonians with  an  ultraviolet
  cutoff}\label{DefHamiUV}

The Fr\"ohlich polaron model descirbes an electron in an ionic crystal
\cite{HFroehlich}.
Despite its long history, this  topic is still being studied actively
\cite{DM, FLST, FLS, GHW, GM, LR, Miyao3}.
The literature on this model is vast  and we content ourselves with mentioning two 
references \cite{Dev, Feynman2}. 
Although the structure of the model is simple, the model has rich  contents both 
mathematically and physically.
For this reason, many researchers employ the model as a touchstone, by
applying their own methods \cite{AL, BFS, DoVa, Feynman, LGross2,LLP,  MS, Pizzo,
Sloan2, Spohn1}.
As a test  of our theory of the operator monotonicity, we will choose this model as well.

\subsubsection{The Fr\"ohlich Hamiltonian}

For each $\Lambda>0$, we define the {\it Fr\"ohlich Hamiltonian}
with an ultraviolet cutoff $\Lambda$
by
\begin{align}
H_{\Lambda}=-\frac{1}{2}\Delta_x-\sqrt{\alpha}\lambda_0
\int_{|k|\le \Lambda}\dm k\, \frac{1}{|k|}\big[\ex^{\im k\cdot x}a(k)+
\ex^{-\im k\cdot x}a(k)
\big]
+\Nf,\label{FullHami}
\end{align}  
where $\Delta_x$ is the  Laplacian on $L^2(\BbbR^3_x)$ and
$\lambda_0=2^{1/4}(2\pi)^{-1}$.
$H_{\Lambda}$ acts in the Hilbert space  $L^2(\BbbR^3_x)\otimes \Fock$.
 Then, by the standard  bound
\begin{align}
\|a(f)^{\#}(\Nf+\one)^{-1/2}\|\le \| f\|\label{StandardBound}
\end{align} 
which comes from (\ref{CreAnnInq}),  and Kato-Rellich theorem \cite{ReSi2}, 
$H_{\Lambda}$ is self-adjoint on $\D(\Delta_x)\cap \D(\Nf)$ and bounded
from below.

\subsubsection{The Fr\"ohlich Hamiltonian at a  fixed total momentum}
Let $P_{\mathrm{tot}}$ be the total momentum operator defined by
\begin{align}
P_{\mathrm{tot}}=-\im \nabla_x+\Pf.
\end{align} 
 Let $\mathcal{F}_x$ be the Fourier
 transformation
on $L^2(\BbbR^3_x)$ and let $U=\mathcal{F}_x \ex^{\im x\cdot \Pf}$. 
 This unitary operator $U$ gives  a
 spectral representation of $P_{\mathrm{tot}}$, namely, 
\begin{align}
U P_{\mathrm{tot}} U^*=\int^{\oplus}_{\BbbR^3} P\, \dm P.
\end{align} 
Moreover  one has
\begin{align}
UH_{\Lambda}U^*=\int^{\oplus}_{\BbbR^3} H_{\Lambda}(P)\, \dm P. \label{Fiber}
\end{align} 
Each $H_{\Lambda}(P)$ is concretely expressed as 
\begin{align}
H_{\Lambda}(P)=\frac{1}{2}(P-\Pf)^2-\sqrt{\alpha}\lambda_0
\int_{|k|\le \Lambda}\dm k\, \frac{1}{|k|}[a(k)+a(k)^*]+\Nf. \label{FixHamiCutt}
\end{align} 
Then, by (\ref{StandardBound}) and Kato-Rellich theorem,
$H_{\Lambda}(P)$ is self-adjoint on $\D(\Pf^2)\cap \D(\Nf)$, bounded
from below.
The self-adjoint operator (\ref{FixHamiCutt}) is called the Fr\"ohlich Hamiltonian
with an ultraviolet cutoff $\Lambda$, at  a fixed total momentum $P$.

\subsection{Results }

\subsubsection{The Fr\"ohlich  Hamiltonian}

To investigate properties of the total Hamiltonian $H_{\Lambda}$, 
it is convenient to move to the electron momentum space:
$L^2(\BbbR_p^3)\otimes \Fock=\mathcal{F}_x L^2(\BbbR^3_x)\otimes \Fock$.
In order to  define the  inequalities  $\unrhd $ etc., we have to fix
 a suitable self-dual cone in $L^2(\BbbR^3_p)\otimes \Fock$.
To this end, let 
\begin{align}
\mathfrak{P}=\big\{
\Psi\in L^2(\BbbR^3_p)\otimes \Fock\, \big|\,
\forall f \in L^2(\BbbR^3_p)_+\ \forall \vphi \in \Fock_+ ,\ 
 \la \Psi, f\otimes
 \vphi\ra\ge 0\ 
\big\}.
\end{align} 
 Then one  sees that $\mathfrak{P}$ is a self-dual cone. Moreover 
$\mathfrak{P}$ can be represented as 
\begin{align}
\mathfrak{P}=\Sumoplus\mathfrak{P}_n
\end{align}
 with $\mathfrak{P}_0=L^2(\BbbR^3_p)_+$ and $\mathfrak{P}_n=
\{
\Psi\in L^2(\BbbR^3_p)\otimes L^2_{\mathrm{sym}}(\BbbR^{3n}_k)
\ |\  \la \Psi, f\otimes \vphi\ra >0\  \forall f\in L^2(\BbbR^3_p)_+
\forall \vphi \in\Fock^{(n)}_+
\}.
$
Remark that, under the identification (\ref{FockIdentify}), one sees
\begin{align}
\mathfrak{P}_n=\{
\Psi\in L^2(\BbbR^3_p)\otimes L_{\mathrm{sym}}^2(\BbbR^{3n}_k)
\ |\  \Psi(p; k_1, \dots, k_n)\ge 0 \ \mbox{a.e.}\}.
\end{align}  

Now let us display our results. Our first result  is as follows.
\begin{Thm}\label{ExistenceFullPolaron}
One has the following.
\begin{itemize}
\item[{\rm (i)}] For any $\Lambda>0$, there exists a constant $M$
independent of $\Lambda$ such that $H_{\Lambda}\ge M$.
\item[{\rm (ii)}] For any $\Lambda>0$, $\mathcal{F}_x \ex^{-t H_{\Lambda}}\mathcal{F}_x^{-1}
	    \unrhd 0$ w.r.t. $\mathfrak{P}$ for all $t\ge 0$, where
	    $\mathcal{F}_x$ is the Fourier transformation associated
	    with $x$.
\item[{\rm (iii)}] $\mathcal{F}_x H_{\Lambda} \mathcal{F}_x^{-1}$ is
	     monotonically decreasing in $\Lambda$ in the sense
\begin{align}
\Lambda' \ge \Lambda \Longrightarrow \mathcal{F}_x H_{\Lambda}\mathcal{F}_x^{-1}\unrhd \mathcal{F}_x
 H_{\Lambda'} \mathcal{F}_x^{-1}\ \ \  w.r.t.\  \mathfrak{P}.
\end{align} 
\end{itemize} 
\end{Thm} 
 A proof of Theorem \ref{ExistenceFullPolaron} will be given in \S 
 \ref{ProofExFullP}.  Now we are ready to apply Theorem \ref{Existence}.
An immediate corollary is as follows.

\begin{coro} \label{FullExistence} 
There exists a self-adjoint operator $H$, bounded from below by
 $M$, with the following properties.
\begin{itemize}
\item[{\rm (i)}] $H_{\Lambda}$ converges to $H$ in strong resolvent sense as
	   $\Lambda\to \infty$.
\item[{\rm (ii)}] For all $\Lambda \ge 0$ and $s\ge 0$, $\mathcal{F}_x \ex^{-s
	    H}\mathcal{F}_x^{-1}\unrhd \mathcal{F}_x \ex^{-s H_{\Lambda}}\mathcal{F}_x^{-1}
$ w.r.t. $\mathfrak{P}$. In particular, $\mathcal{F}_x\ex^{-s H}
\mathcal{F}_x^{-1}\unrhd 0$
w.r.t. $\mathfrak{P}$ for all $s\ge 0$. 
\end{itemize} 
\end{coro} 
\begin{rem}
{\rm
There are several ways to define the Hamiltonian $H$ as a limiting
 operator \cite{GlimmJaffe,LGross2,  Nelson, Sloan2}.
Here we propose a novel method by the operator monotonicity.
To keep a decent form of the article, we exhibit results here, however 
the readers should pay attention to our proof. $\diamondsuit$
}
\end{rem} 
We will give a proof of Corollary \ref{FullExistence} in \S \ref{CoroFullExistence}.
By the above corollary, we can define the Fr\"ohlich Hamiltonian
without ultraviolet cutoff by $H$.

\subsubsection{The Fr\"ohlich  Hamiltonian   at a fixed total momentum}
Next we will state  results on $H_{\Lambda}(P)$.

\begin{Thm}\label{MonotoneFixHami}
One has the following.
\begin{itemize}
\item[{\rm (i)}] For any $\Lambda>0$, there exists a constant $M$
independent of $\Lambda$ and $P$ such that $H_{\Lambda}(P)\ge M$.
\item[{\rm (ii)}] For any $\Lambda>0$, $ \ex^{-t H_{\Lambda}(P)}
	    \unrhd 0$ w.r.t. $\Fock_+$ for all $t\ge0$ and $P\in \BbbR^3$.
\item[{\rm (iii)}]  For all $P\in \BbbR^3$, $H_{\Lambda}(P)$ is
	     monotonically decreasing in $\Lambda$ in the sense 
\begin{align}
\Lambda' \ge \Lambda \Longrightarrow H_{\Lambda}(P) \unrhd 
 H_{\Lambda'}(P)\ \ \  w.r.t.\  \Fock_+. \label{FixHMonotonicity}
\end{align} 

\end{itemize} 
\end{Thm}

We will show Theorem \ref{MonotoneFixHami} in \S \ref{ProofMonoFixH}.
Applying Theorem \ref{Existence}, one has the following corollary.

\begin{coro} \label{ExsPolaronFix}
 There exists a self-adjoint operator $H(P)$, bounded from below by
 $M$, with the following properties.
\begin{itemize}
\item[{\rm (i)}] $H_{\Lambda}(P)$ converges to $H(P)$ in strong resolvent sense as
	   $\Lambda\to \infty$.
\item[{\rm (ii)}] For all $\Lambda\ge 0,\  P\in \BbbR^3$ and $s\ge 0$, $ \ex^{-s
	    H(P)}\unrhd  \ex^{-s H_{\Lambda}(P)}
$ w.r.t. $\Fock_+$. In particular, $\ex^{-s H(P)}\unrhd 0$
w.r.t. $\Fock_+$ for all $P\in \BbbR^3$ and $s\ge 0$. 
\item[{\rm (iii)}] Let $H$ be  the Hamiltonian in Theorem
	    \ref{FullExistence}. Then one has 
\begin{align}
UHU^{-1}=\int_{\BbbR^3}^{\oplus} H(P)\, \dm P.
\end{align} 
\end{itemize} 
\end{coro} 

We will give a proof of Corollary \ref{ExsPolaronFix} in \S \ref{ProofExsPFix}.
In this way, we can define the Hamiltonian without ultraviolet cutoff by $H(P)$.

As to the limiting Hamiltonian  $H(P)$, we can say more.

\begin{Thm}\label{PosiImpPolaron}
 Let $H(P)$ be  the self-adjoint operator in Theorem \ref{ExsPolaronFix}.
Then one has  
\begin{align}
\ex^{-s H(P)} \rhd 0\
\end{align} 
 w.r.t. $\Fock_+$ for all $P\in \BbbR^3$ and  $s>0$. Consequently, if $\inf \mathrm{spec}(H(P))$
 is an eigenvalue, then it is unique and the corresponding eigenvector 
 can be chosen as  strictly positive w.r.t. $\Fock_+$.
\end{Thm} 

We will prove Theorem \ref{PosiImpPolaron} in \S \ref{ProofPIP}.
Existence of a ground state of $H(P)$ was fully understood, namely, it
was already shown $H(P)$ has a ground state provided $|P|<\sqrt{2}$
\cite{GeLowen, Spohn2}. Combining this and Theorem \ref{Faris}, we arrive at the
corollary below.

\begin{coro}
For all $P\in \BbbR^3$ with $|P|<\sqrt{2}$, $H(P)$ has a unique ground
 state which is strictly positive w.r.t. $\Fock_+$. 
\end{coro}

\subsubsection{Monotonicity of the polaron energy \cite{Miyao3}}

By the operator monotonicity (\ref{FixHMonotonicity}), we can further
obtain information about the ground state energy of $H_{\Lambda}(P)$.
Here we only exhibit the results proved in \cite{Miyao3}.

\begin{Thm}{\rm \cite{Miyao3}}\label{EnergyMonoPolaron}
Under assumptions in Theorem \ref{ExsPolaronFix},
set $E_{\Lambda}(P)=\inf \mathrm{spec}(H_{\Lambda}(P))$
and $E(P)=\inf \mathrm{spec}(H(P))$. 
 Then, for all $|P|<\sqrt{2}$, 
$E_{\Lambda}(P)$ is strictly decreasing in $\Lambda$.
\end{Thm} 

Let $E_{\Lambda}=\inf \mathrm{spec}(H_{\Lambda})$. Then, by
(\ref{Fiber}) and the fact $E_{\Lambda}(0) \le E_{\Lambda}(P)$, 
one has $E_{\Lambda}=E_{\Lambda}(0)$. This equality implies the
following.

\begin{coro}Under assumptions in Theorem \ref{ExsPolaronFix},
set $E=\inf \mathrm{spec}(H)$.  Then 
$E_{\Lambda}$ is strictly decreasing in $\Lambda$.
\end{coro} 

In this way, our method of the operator monotonicity gives a consistent
theory  of the Fr\"ohlich polaron.

\section{Proof of Theorem \ref{ExistenceFullPolaron}}\label{ProofExFullP}
\setcounter{equation}{0}

\subsection{Proof of (i): Uniform lower bound}

\begin{Prop} \label{LiebYamazaki}
Take  $0<\Lambda_0 <\infty$ such that 
\begin{align}
\alpha \lambda^2_0\int_{|k|\ge \Lambda_0}\dm k\, \frac{1}{|k|^4}<\frac{1}{8}.\label{Integral}
\end{align} 
Then, for any $\Lambda$ satisfying $\Lambda \ge \Lambda_0$, one has 
\begin{align*}
H_{\Lambda}\ge -\alpha \lambda_0^2 
\int_{|k|\le \Lambda_0}\dm k\, \frac{1}{|k|^2}-\frac{1}{2}.
\end{align*} 
\end{Prop}
{\it Proof.}
We will employ a   method established by Lieb-Yamazaki \cite{LY}. (See
also \cite{LT1}.) 
Pick $\Lambda$ such that $\Lambda \ge \Lambda_0$.
Let $\chi_{\Lambda}(k)=1$ if $|k|\le \Lambda$,
$\chi_{\Lambda}(k)=0$ otherwise.
Define $\mathbf{Z}=(Z_1, Z_2, Z_3)$ by 
\begin{align}
Z_j=\sqrt{\alpha}\lambda_0\int_{\BbbR^3_k}\dm k\, k_j
\frac{\chi_{\Lambda}(k)-\chi_{\Lambda_0}(k)}{|k|^3} \ex^{\im k\cdot x} a(k).
\end{align} 
Note that $k_j(\chi_{\Lambda}(k)-\chi_{\Lambda_0}(k))/|k|^3$ is
square-integrable. 
Set
\begin{align}
I(f)=\sqrt{\alpha}\lambda_0\int_{\BbbR^3_k} \dm k\, \frac{f(k)}{|k|}[\ex^{\im k\cdot
 x}a(k)
+\ex^{-\im k\cdot x}a(k)^*
].
\end{align}
Then, using  basic  operator inequalities
$\displaystyle 
2^{-1}\varepsilon a^*a+2\vepsilon^{-1} b^*b\ge \pm (a^*b+b^*a)
$
for any $\vepsilon>0$, one sees
\begin{align}
I(\chi_{\Lambda}-\chi_{\Lambda_0})=&\sum_{j=1,2,3}[-\im \nabla_x, Z_j-Z_j^*]\no
\le&\vepsilon (-\Delta_x)+2\vepsilon^{-1}(\mathbf{Z}^*\cdot
 \mathbf{Z}+\mathbf{Z}\cdot \mathbf{Z}^*)\no
=& \vepsilon (-\Delta_x)+4\vepsilon^{-1}\mathbf{Z}^*\cdot
 \mathbf{Z}+2\vepsilon^{-1}\alpha \lambda^2_0 \int \dm k\,
 \frac{(\chi_{\Lambda}(k)-\chi_{\Lambda_0}(k))^2}{|k|^4}\ \ \ \mbox{(By  CCRs)}\no
\le& \vepsilon (-\Delta_x) +4\vepsilon^{-1}\alpha \lambda_0^2
\int \dm k
 \frac{\chi_{\Lambda}(k)-\chi_{\Lambda_0}(k)}{|k|^4}\dG(\chi_{\Lambda }-\chi_{\Lambda_0})
\no
&+2\vepsilon^{-1}\alpha \lambda^2_0 \int \dm k\,
 \frac{\chi_{\Lambda}(k)-\chi_{\Lambda_0}(k)}{|k|^4}.\ \ \ \ \ \mbox{(By (\ref{CreAnnInq}))}\label{IneqI}
\end{align} 
Take $\vepsilon=4\alpha \lambda_0^2\int \dm k\,
(\chi_{\Lambda}(k)-\chi_{\Lambda_0}(k))/|k|^4$.  Then (\ref{IneqI}) becomes 
\begin{align}
I(\chi_{\Lambda}-\chi_{\Lambda_0})\le \vepsilon(-\Delta_x)+\dG(\one-\chi_{\Lambda_0})+\frac{1}{2}.
\end{align} 
Hence, by (\ref{VHove}),
\begin{align}
H_{\Lambda}&= H_{\Lambda_0}-I(\chi_{\Lambda}-\chi_{\Lambda_0})\no
&\ge \frac{1}{2}(1-2\vepsilon)
 (-\Delta_x)-I(\chi_{\Lambda_0})+\dG(\chi_{\Lambda_0})-\frac{1}{2}\no
&\ge \dG(\chi_{\Lambda_0})-I(\chi_{\Lambda_0})-\frac{1}{2}\no
&\ge -\alpha \lambda_0^2 \int\dm k\,
 \frac{\chi_{\Lambda_0}(k)}{|k|^2}-\frac{1}{2}.
\end{align} 
This proves the assertion. $\Box$
\medskip\\

Fix $\Lambda_0$ arbitarily so that (\ref{Integral}) holds. Then, for
each $\Lambda  \ge \Lambda_0$, we have 
\begin{align}
H_{\Lambda} \ge -\alpha \lambda_0^2 \int_{|k|\le \Lambda_0} \dm k
 \frac{1}{|k|}-\frac{1}{2} \label{LE1}
\end{align} 
by Proposition \ref{LiebYamazaki}.
On the other hand, for $0 \le \Lambda\le \Lambda_0 $, one sees
\begin{align}
H_{\Lambda} &\ge \Nf-\sqrt{\alpha} \lambda_0 \int_{|k|\le \Lambda}\dm k
\frac{1}{|k|}[\ex^{\im k\cdot x} a(k)+\ex^{-\im k\cdot x} a(k)^*]\no
&\ge -\alpha \lambda_0^2\int_{|k|\le \Lambda}\dm k\frac{1}{|k|^2}\label{LE2}
\end{align} 
by (\ref{VHove}).
Combining (\ref{LE1}) with (\ref{LE2}), one arrives at 
\begin{align}
H_{\Lambda} \ge  -\alpha \lambda_0^2 \int_{|k|\le \Lambda_0} \dm k
 \frac{1}{|k|}-\frac{1}{2} \label{LE3}
\end{align} 
for {\it any} $\Lambda\ge 0$. $\Box$

\subsection{Proof of (ii): Positivity preserving property}
Let us denote  a transformed  Hamiltonian $\mathcal{F}_x
H_{\Lambda} \mathcal{F}_x^{-1}$ by $\hat{H}_{\Lambda}$.
$\hat{H}_{\Lambda}$ acts in the Hilbert space 
 $L^2(\BbbR^3_p)\otimes
\Fock=\mathcal{F}_x L^2(\BbbR^3_x)\otimes \Fock$. In addition,
$\hat{H}_{\Lambda}$ 
has the following form
\begin{align}
\hat{H}_{\Lambda}=\mathcal{F}_xH_{\Lambda} \mathcal{F}_x^{-1}
=\hat{H_0}-\hat{V}_{\Lambda}\no
\end{align} 
with 
\begin{align}
\hat{H}_0&=\frac{1}{2}p^2+\Nf,\\
\hat{V}_{\Lambda}&=\sqrt{\alpha}\lambda_0\int_{|k|\le \Lambda}\dm k\, 
\frac{1}{|k|}\big[\ex^{\im k\cdot (-\im \nabla_p)}
a(k)+\ex^{-\im k\cdot (-\im \nabla_p)}a(k)^*
\big],\label{ElectronMom}
\end{align} 
where $p^2$ is a multiplication operator and $\nabla_p$ is the standard
nabla symbol on $L^2(\BbbR^3_p)$.
As we will see, the expression (\ref{ElectronMom}) is essential
for our proof.

\begin{lemm}{\rm (Attraction)} For any $\Lambda \ge 0$, $-V_{\Lambda}$ is
 attractive in the sense $-V_{\Lambda} \unlhd 0$
w.r.t. $\mathfrak{P}$.
\end{lemm} 
{\it Proof.}
Recall the expression (\ref{ElectronMom}).
Since $\ex^{\im k\cdot (-\im \nabla_p)}$ is a translation,  it
satisfies $\ex^{\im k\cdot (-\im \nabla_p)}\unrhd 0$
w.r.t. $L^2(\BbbR^3_p)_+$.
Hence one also concludes, for any $\Lambda \ge 0$, 
\begin{align}
\int_{|k|\le \Lambda}\dm k\, \frac{1}{|k|}\ex^{\im k\cdot (-\im
\nabla_p)}a(k)\unrhd 0
\end{align} 
 w.r.t. $\mathfrak{P}$.
Accordingly its adjoint operator is also positivity preserving.
Now we conclude the assertion in the lemma. $\Box$
\medskip\\

Since $\ex^{-s \one} \unrhd 0$
w.r.t. $L^2(\BbbR^3_k)_+$,
one has, by Proposition \ref{PPFockI},  $\ex^{-s \Nf}=\Gamma(\ex^{-s \one})\unrhd 0$ w.r.t. $\Fock_+$ which
implies $\ex^{-s (\frac{1}{2}p^2+\Nf)}=\ex^{-s\frac{1}{2}p^2}\ex^{-s
\Nf}\unrhd 0$ w.r.t. $ \mathfrak{P}$.
Here we used the fact the multiplication operator $\ex^{-s
\frac{1}{2}p^2}$
preserves the positivity w.r.t. $\mathfrak{P}$.
Then we can apply Corollary  \ref{PerturbationPP} and  conclude 
$\ex^{-s H_{\Lambda}}\unrhd 0$ w.r.t. $\mathfrak{P}$.  This completes
the proof of (ii).
$\Box$

\subsection{Proof of (iii): Operator monotonicity}
Let us begin with the following lemma.
\begin{lemm}\label{VMono}
$-\hat{V}_{\Lambda}$ is monotonically decreasing in $\Lambda$ in a sense that 
\begin{align}
\Lambda_1\le \Lambda_2 \Longrightarrow -\hat{V}_{\Lambda_1} \unrhd
 -\hat{V}_{\Lambda_2}\ \ \ \mbox{w.r.t. $\mathfrak{P}$}.
\end{align} 
\end{lemm} 
\begin{rem}{\rm
The attraction becomes stronger, the larger  we take the ultraviolet
 cutoff. This is the physical meaning of the above lemma.
}
\end{rem} 
{\it Proof.}
Suppse $\Lambda_1 \le \Lambda_2$.
Then, for each  $\vphi\in \D(\hat{V}_{\Lambda_1})\cap \D(\hat{V}_{\Lambda_2})$,
one has 
\begin{align}
(\hat{V}_{\Lambda_2}-\hat{V}_{\Lambda_1})\vphi =\sqrt{\alpha}\lambda_0
 \int_{\BbbR^3}
\dm k\, \frac{\chi_{\Lambda_2}(k)-\chi_{\Lambda_1}(k)}{|k|}
\big[\ex^{\im k\cdot (-\im \nabla_p)}
a(k)+\ex^{-\im k\cdot (-\im \nabla_p)}a(k)^*
\big]\vphi. \label{DiffV}
\end{align} 
If $\vphi\ge 0$ w.r.t. $\mathfrak{P}$, then the right hand side of
(\ref{DiffV}) is also positive w.r.t. $\mathfrak{P}$.
 Hence one concludes the assertion. $\Box$
\medskip\\

Suppose $\Lambda \le \Lambda'$.
For any $\vphi\in \D(p^2)\cap \D(\Nf)=\D(\hat{H}_{\Lambda})=\D(\hat{H}_{\Lambda'})$,
\begin{align}
(\hat{H}_{\Lambda}-\hat{H}_{\Lambda'})\vphi= (\hat{V}_{\Lambda'}-\hat{V}_{\Lambda})\vphi.
\label{Difference}
\end{align} 
Hence if $\vphi\ge 0$ w.r.t. $\mathfrak{P}$,  the right hand side of
(\ref{Difference}) is positive w.r.t. $\mathfrak{P}$ by Lemma \ref{VMono}.
Hence one concludes $\hat{H}_{\Lambda}\unrhd \hat{H}_{\Lambda'}$. 
w.r.t. $\mathfrak{P}$.
$\Box$

\section{Proof of Corollary \ref{FullExistence}}\label{CoroFullExistence}

By Theorem \ref{ExistenceFullPolaron},  $\mathcal{F}_x
H_{\Lambda}\mathcal{F}_x^{-1}
$ is monotonically decreasing in $\Lambda$ and uniformly bounded from
below. 
Moreover $\ex^{-t \mathcal{F}_x H_{\Lambda} \mathcal{F}_x^{-1}}\unrhd 0$
w.r.t. $\mathfrak{P}$ holds.
This means the assumptions  {\bf (A. 1)},  {\bf (A. 2)} and (\ref{Monotonicity}) in Theorem 
\ref{Existence} are satisfied.
 Moreover each $\hat{H}_{\Lambda}(P)$ has the  common domain
 $\mathscr{D}=\D(p^2)\cap \D(\Nf)$, hence the assumption {\bf (A. 3)}
is satisfied as well. 
Now we
can apply Theorem \ref{Existence} and conclude (i) and (ii).
 $\Box$

\section{Proof of Theorem \ref{MonotoneFixHami}}\label{ProofMonoFixH}
\setcounter{equation}{0}

\subsection{Proof of (i): Uniform lower bound}
Choose $0 \le \Lambda_0 <\infty$ such that $\sqrt{\alpha} \lambda_0 \int \dm k\,
(1-\chi_{\Lambda_0}(k))/|k|^4<1/8$. Then, by (\ref{LE3}),
for any $\Lambda\ge0$,  one has a uniform lower bound
\begin{align}
H_{\Lambda} \ge M_0 \label{ULowerBd}
\end{align} 
with $M_0=-\alpha \lambda_0^2 \int \dm k\, \chi_{\Lambda_0}(k)/|k|^2-1/2$.
On the other hand, by the similar arguments  in \S \ref{DefHamiUV}, 
 we have
\begin{align}
UH_{\Lambda}U^{-1}=\int_{\BbbR^3}^{\oplus} H_{\Lambda}(P)\, \dm P
\end{align} 
with $U=\mathcal{F}_x \ex^{\im x\cdot \Pf}$. Hence $\inf_{P\in \BbbR^3}
E_{\Lambda}(P)=E_{\Lambda}$
which implies $\inf E_{\Lambda}(P)\ge E_{\Lambda}$ for almost every $P$. But since $E_{\Lambda}(P)$ is continuous in  $P$, this inequality
holds true for {\it all} $P\in \BbbR^3$. [Proof of the continuity:
For all $P$, $H_{\Lambda}(P)$ has the  common domain
$\mathcal{D}=\D(\Pf^2)\cap \D(\Nf)$.
Then, for all $\vphi\in \mathcal{D}$, one easily sees
$H_{\Lambda}(P')\vphi\to H_{\Lambda}(P)\vphi$ as $P'\to P$ which implies 
$H_{\Lambda}(P')$ converges to $H_{\Lambda}(P)$ in strong resolvent
sense
by \cite[Theorem VIII. 25]{ReSi2}.
]
Combining this with (\ref{ULowerBd}), we have $H_{\Lambda}(P)\ge M_0$
 for all $P\in \BbbR^3$. $\Box$

\subsection{Proof of (ii): Positivity preserving property}

We express the Hamiltonian $H_{\Lambda}(P)$ as 
\begin{align}
H_{\Lambda}(P)=H_0(P)-W_{\Lambda}
\end{align} 
with 
\begin{align}
H_0(P)&= \frac{1}{2}(P-\Pf)^2+\Nf,\\
W_{\Lambda}&= \sqrt{\alpha}\lambda_0 \int_{|k|\le \Lambda}\dm k\, 
 \frac{1}{|k|} [a(k)+a(k)^*].
\end{align} 
\begin{lemm}{\rm  (Attraction)} For any $\Lambda \ge 0$, $-W_{\Lambda}$ is
 attractive in a sense $-W_{\Lambda} \unlhd 0$ w.r.t. $\Fock_+$.
\end{lemm} 
{\it Proof.}
For any $\Lambda \ge 0$, 
we have $\sqrt{\alpha}\lambda_0 \int \dm k\,
\chi_{\Lambda}(k)a(k)/|k|\unrhd 0$ w.r.t. $\Fock_+$ by Proposition
\ref{PPFockII}.
Of course its adjoint also preserves the positivity w.r.t. $\Fock_+$.
Thus we conclude the assertion in the lemma. $\Box$
\medskip\\

By Proposition \ref{PPFockI}, $\ex^{-t \Nf}\unrhd 0$ w.r.t. $\Fock_+$.
Furthermore $\ex^{-t (P-\Pf)^2}\unrhd 0$ w.r.t. $\Fock$ for all $P$.
[Proof:
We can write $\ex^{-t (P-\Pf)^2}=\ex^{-t |P|^2} \oplus \sum_{n\ge 1}^{\oplus}
\exp[-t (P-\sum_{j=1}^n k_j)^2]$. Each $n$-th component satisfies $\exp[-t (P-\sum_{j=1}^n k_j)^2] \unrhd 0$
w.r.t. $\Fock_+^{(n)}$.] This implies $\exp[-tH_0(P)]
=\exp[-t \frac{1}{2}(P-\Pf)^2]\exp[-t \Nf] \unrhd 0$ w.r.t. $\Fock_+$
for all $P$. 
 Now we can apply Corollary \ref{PerturbationPP} and  conclude $\ex^{-t
H_{\Lambda}(P)}\unrhd 0$ w.r.t. $\Fock_+$.  
$\Box$

\subsection{Proof of (iii): Operator monotonicity}
First of all, we clarify the monotonicity of the interaction term:
\begin{lemm}
\label{WMono}
$-W_{\Lambda}$ is monotonically decreasing in $\Lambda$ in a sense that 
\begin{align}
\Lambda_1\le \Lambda_2 \Longrightarrow -W_{\Lambda_1} \unrhd
 -W_{\Lambda_2}\ \ \ \mbox{w.r.t. $\Fock_+$}.
\end{align} 
\end{lemm}
\begin{rem}{\rm
As before, the attraction becomes stronger, the larger  we take the ultraviolet
 cutoff. 
}
\end{rem} 
 {\it Proof.}
 Suppose that 
$\Lambda_1 \le \Lambda_2$. Then, for any $\vphi\in \D(W_{\Lambda_1})
 \cap \D(W_{\Lambda_2})$, one has 
\begin{align}
(W_{\Lambda_2}-W_{\Lambda_1})\vphi=\sqrt{\alpha} \lambda_0
\int_{\BbbR^3_k}\dm k\,
 \frac{\chi_{\Lambda_2}(k)-\chi_{\Lambda_1}(k)}{|k|}[a(k)+a(k)^*]
 \vphi. \label{DeffFix}
\end{align} 
Hence if $\vphi \ge 0$ w.r.t. $\Fock_+$, then the right hand side of
(\ref{DeffFix}) is positive w.r.t. $\Fock_+$ as well.
Hence we obtain the assertion in the lemma. $\Box$
\medskip\\

Suppose $\Lambda\le \Lambda'$. 
For any $\vphi\in \D(\Pf^2)\cap
\D(\Nf)=\D(H_{\Lambda})=\D(H_{\Lambda'})$,
we have 
\begin{align}
(H_{\Lambda}-H_{\Lambda'})\vphi =(W_{\Lambda'}-W_{\Lambda})\vphi.\label{DiffW}
\end{align} 
Thus if $\vphi\ge 0$ w.r.t. $\Fock_+$, then 
 the right hand side of (\ref{DiffW}) is also  positive
w.r.t. $\Fock_+$ by Lemma \ref{WMono}. This ompletes the proof of (iii).
 $\Box$

\section{Proof of Corollary \ref{ExsPolaronFix}}\label{ProofExsPFix}

\subsection{Proof of (i) and (ii): Existence of the limit}
By Theorem \ref{MonotoneFixHami},  $H_{\Lambda}(P)$
is monotonically  decreasing in $\Lambda$, uniformly bounded from below, and $\ex^{-t H_{\Lambda}(P)} \unrhd
0$ w.r.t. $\Fock_+$ for all $\Lambda \ge  0, P\in \BbbR^3$ and $t\ge 0$.
Moreover $H_{\Lambda}(P)$ has the common domain
$\mathscr{D}=\D(\Pf^2)\cap \D(\Nf)$. Thus the assumptions {\bf (A. 1)},
{\bf (A. 2)}, {\bf (A. 3)} and (\ref{Monotonicity}) are satisfied. 
Hence we can apply Theorem \ref{Existence} and  conclude the existence of
the limit $H(P)$ satisfying (i) and (ii). $\Box$

\subsection{Proof of (iii): Decomposition of the limit  $H$}
For each $\Lambda \ge 0$, one has 
$U H_{\Lambda}U^{-1}=\int^{\oplus}_{\BbbR^3} H_{\Lambda}(P)\, \dm
P$. Thus 
\begin{align}
U\ex^{-t H_{\Lambda}}U^{-1}=\int^{\oplus}_{\BbbR^3} \ex^{-t
 H_{\Lambda}(P)}
\, \dm P.
\end{align}
Taking $\Lambda\to \infty$, one arrives at 
\begin{align}
U\ex^{-t H}U^{-1}=\int^{\oplus}_{\BbbR^3} \ex^{-t
 H(P)}
\, \dm P
\end{align}
which implies the desired assertion. $\Box$

\section{Proof of Theorem \ref{PosiImpPolaron}} \label{ProofPIP}
\setcounter{equation}{0}
In the previous work \cite{Miyao}, the author proved $\ex^{-t
H(P)}\rhd 0$ w.r.t. $\Fock_+$  for all $t>0$ by applying Theorem
\ref{PositivityImp}. In that proof, a mild ultraviolet cutoff was
employed.
In this paper, we are treating the sharp cutoff $\chi_{\Lambda}$, so
that the
method in \cite{Miyao} can not be applied directly. In this section, we
will provide an alternative  proof based on the local ergodicity. This
method
can cover the case of the sharp ultraviolet cutoff.

To prove Theorem \ref{PosiImpPolaron}, we introduce a {\it local
Hamiltonian} $K_{\Lambda}(P)$ by 
\begin{align}
K_{\Lambda}(P)=K_0(P)-W_{\Lambda}
\end{align} 
with 
\begin{align}
K_0(P)&=\frac{1}{2}(P-P_{\mathrm{f}, \Lambda})^2+N_{\mathrm{f},
 \Lambda},\\
W_{\Lambda}&= \sqrt{\alpha} \lambda_0 \int_{|k|\le \Lambda}\dm k\,
 \frac{1}{|k|}[a(k)+a(k)^*].
\end{align} 
Here 
\begin{align}
P_{\mathrm{f}, \Lambda}=\int_{|k|\le \Lambda}\dm k\, k a(k)^*a(k),\ \ \
 N_{\mathrm{f}, \Lambda}=\int_{|k|\le \Lambda}\dm k\, a(k)^*a(k).
\end{align} 
The following local property is essential for our study.
\begin{lemm}\label{LocalHamiAndH}
Choose $\mu>0$ such that $H_{\Lambda}+\mu>0$ for all $\Lambda \ge 0$.
Then, for all $0\ \le \Lambda <\infty$ and $P\in \BbbR^3$, one obtains 
\begin{align}
Q_{\Lambda}(H_{\Lambda}(P)+\mu)^{-1}Q_{\Lambda}=Q_{\Lambda}(K_{\Lambda}(P)+\mu)^{-1}Q_{\Lambda}.
\end{align} 
\end{lemm} 
{\it Proof.} First we remark that, for each $f\in L^2(B_{\Lambda})$,
$Q_{\Lambda} a(f)=a(f)Q_{\Lambda}$ and $Q_{\Lambda}a(f)^*=a(f)^*
Q_{\Lambda}$ hold.
Furthermore since $\Pf Q_{\Lambda}=P_{\mathrm{f},
\Lambda}Q_{\Lambda}=Q_{\Lambda} P_{\mathrm{f}, \Lambda}$ and 
$\Nf Q_{\Lambda}=N_{\mathrm{f},
\Lambda}Q_{\Lambda}=Q_{\Lambda}N_{\mathrm{f}, \Lambda}$, one sees
\begin{align}
H_{\Lambda}(P)Q_{\Lambda}=Q_{\Lambda}K_{\Lambda}(P).
\end{align} 
Thus 
\begin{align}
Q_{\Lambda}(K_{\Lambda}(P)+\mu)^{-1}=(H_{\Lambda}(P)+\mu)^{-1}Q_{\Lambda}
\end{align} 
holds. This completes the proof. $\Box$
\medskip\\

The following proposition was  proved in the previous work \cite{Miyao}.

\begin{Prop}\label{LocalErgodicity}{\rm (Local ergodicity)}
For all $\Lambda>0, P\in \BbbR^3$ and $t>0$, one obtains 
$
\ex^{-t K_{\Lambda}(P)}\rhd 0
$
w.r.t. $\Fock_{\Lambda, +}$.
\end{Prop} 
{\it Proof of Theorem \ref{PosiImpPolaron}}\\
Choose $\vphi, \psi\in \Fock_+\backslash \{0\}$ arbitrarily.
Since $\displaystyle \mbox{s-}\lim_{\Lambda\to \infty}Q_{\Lambda}=\one$, 
there exists  a $\Lambda$ such that $Q_{\Lambda}\vphi\neq 0$ and
$Q_{\Lambda}\psi \neq 0$.  This means $Q_{\Lambda}\vphi,
Q_{\Lambda}\psi\in \Fock_{\Lambda, +}\backslash \{0\}.  $ Moreover since $Q_{\Lambda} \unrhd 0$ and
$Q^{\perp}_{\Lambda}=\one -Q_{\Lambda}\unrhd 0$ w.r.t. $\Fock_+$ by
Propositon \ref{QPP}, one
has 
$\vphi \ge Q_{\Lambda}\vphi$ and $\psi \ge Q_{\Lambda }\psi$
w.r.t. $\Fock_+$. 
Note, for $\mu$ sufficiently large, $(K_{\Lambda}(P)+\mu)^{-1}\rhd 0$
w.r.t. $\Fock_{\Lambda, +}$ by Proposition \ref{LocalErgodicity} and
Theorem \ref{Faris}.
Hence  one has, by  Lemma \ref{LocalHamiAndH}, 
\begin{align}
\la \psi, (H_{\Lambda}(P)+\mu)^{-1}\vphi \ra &\ge \la
 Q_{\Lambda}\psi, (H_{\Lambda}(P)+\mu)^{-1}Q_{\Lambda}\vphi\ra_{\Fock}\no
&= \la
 Q_{\Lambda}\psi, (K_{\Lambda}(P)+\mu)^{-1}Q_{\Lambda}\vphi\ra_{\Fock_{\Lambda}}\ \
 \ \mbox{(By Lemma \ref{LocalHamiAndH})}\no
&>0. 
\end{align} 
Therefore $\{(H_{\Lambda}(P)+\mu)^{-1}\}_{\Lambda}$ is ergodic. Now we can
apply Theorem \ref{Ergodic} and conclude that $\ex^{-s H(P)}\rhd 0$
w.r.t. $\Fock_+$. $\Box$

\appendix

\section{Preliminaries  }
\setcounter{equation}{0}
In this section, we will review some preliminary results about the
operator inequalities introduced in \S \ref{Def}. Almost all of results here are
taken from the author's previous work \cite{Miyao, Miyao2, Miyao3,
Miyao4, Miyao5}.

\subsection{Operator monotonicity}

\begin{Prop}{\rm (Monotonicity)}\label{ABIneqEq}
Let $A$ and $B$ be positive self-adjoint operators. We assume the following.
\begin{itemize}
\item[{\rm (a)}] $\D(A)\subseteq\D(B)$ or $\D(A)\supseteq \D(B)$.
\item[{\rm (b)}] $(A+s)^{-1}\unrhd 0$  and  $(B+s)^{-1}\unrhd 0$ w.r.t. $\Cone$ for all $s>0$.
\end{itemize} 
Then the following are equivalent to each other.
\begin{itemize}
 \item[{\rm (i)}] $B\unrhd A$ w.r.t. $\Cone$.
 \item[{\rm (ii)}] $(A+s)^{-1}\unrhd (B+s)^{-1}$ w.r.t. $\Cone$ for all $s>0$.
 \item[{\rm (iii)}] $\ex^{-tA}\unrhd \ex^{-tB}$ w.r.t. $\Cone$ for all $t\ge 0$.
 \end{itemize}
\end{Prop}
{\it Proof.} See \cite{Miyao, Miyao2}. $\Box$

\begin{coro}\label{PerturbationPP}
Let $A$ be a positive self-adjoint operator and let $B$ be a  symmetric
 operator. Assume the following.
\begin{itemize}
\item[{\rm (i)}] $B$ is $A$-bounded with relative bound $a<1$, i.e.,
                 $\D(A)\subseteq \D(B)$ and $\|Bx\|\le a \|Ax\|+b\|x\|$
                 for all $x\in \D(A)$.
\item[{\rm (ii)}] $0\unlhd \ex^{-tA}$ w.r.t. $\Cone$ for all $t\ge 0$.
\item[{\rm (iii)}]$0\unlhd -B$ w.r.t. $\Cone$.
\end{itemize} 
Then $\ex^{-t(A+B)}\unrhd \ex^{- tA}\unrhd 0$ w.r.t. $\Cone$ for all $t\ge 0$.
\end{coro} 
{\it Proof.} See \cite{Miyao, Miyao2}. $\Box$

\subsection{Perron-Frobenius-Faris theorem}

\begin{Thm}\label{Faris}{\rm (Perron-Frobenius-Faris)}
Let $A$ be a positive self-adjoint operator on $\h$. Suppose that 
 $0\unlhd \ex^{-tA}$ w.r.t. $\Cone$ for all $t\ge 0$ and $\inf
 \mathrm{spec}(A)$ is an eigenvalue.
Let $P_A$ be the orthogonal projection onto the closed subspace spanned
 by  eigenvectors associated with   $\inf
 \mathrm{spec}(A)$.
 Then the following
 are equivalent.
\begin{itemize}
\item[{\rm (i)}] 
$\dim \ran P_A=1$ and $P_A\rhd 0$ w.r.t. $\Cone$.
\item[{\rm (ii)}] $0\lhd (A+s)^{-1}$ for some $s>0$.
\item[{\rm (iii)}] For all $x,y\in\Cone\backslash \{0\}$, there exists a 
  $t> 0$ such that 
$0<\la x,\ex^{-tA}y\ra$.
\item[{\rm (iv)}] $0\lhd (A+s)^{-1}$ for all $s>0$.
\item[{\rm (v)}] $0\lhd \ex^{-tA}$ for all $t>0$.
\end{itemize}
\end{Thm} 
{\it Proof.} See, e.g., \cite{Faris, Miyao, ReSi4}. $\Box$

\section{A remark on the strongly continuous semigroup}
A family of operators $\{T_s\, |\, 0\le s < \infty\}$ on a Hilbert space
$\h$ is called a {\it one-parameter semigroup} if 
\begin{itemize}
\item[{\rm (a)}] $T_0=\one$,
\item[{\rm (b)}] $T_s T_t=T_{s+t}$ for all $s, t \ge 0$.
\end{itemize} 
 In addition if $T_s$ satisfies
\begin{itemize}
\item[{\rm (c)}] for each $x\in \h$, $s\mapsto T_s x$ is strongly continuous,
\end{itemize} 
then the family is called a {\it strongly continuous semigroup}.

The following lemma is well-known \cite{ReSi2}. 
\begin{lemm}\label{SContSemiG}
Let $T_s$ be  a strongly continuous semigroup on a Hilbert space $\h$
 and $\displaystyle Ax=\lim_{s\to +0} \frac{1}{s}(\one-T_s)x$, where
\begin{align}
\D(A)=\Big\{x\in \h\, \Big|\, \lim_{s\to +0}\frac{1}{s}(\one-T_s)x\ \ \mbox{exists}
\Big\}.
\end{align} 
Then $A$ is closed and densely defined.
\end{lemm} 

If we further add conditions on $T_s$, we can prove the self-adjointness
of $A$ as follows.

\begin{Prop}\label{SemiG}
Let $T_s$ be a strongly continuous semigroup on a Hilbert space
 $\h$. Assume 
\begin{itemize}
\item[{\rm (d)}] $T_s$ is self-adjoint for all $s\ge 0$.
\item[{\rm (e)}] There exists an $M>0$ such that $\|T_s\| \le \ex^{-s
	     M}$ for all  $s\ge 0$.
\end{itemize} 
Then $A$ is self-adjoint, bounded from below by $M$ and $T_s=\ex^{-s A}$.
\end{Prop} 
{\it Proof.} By Lemma \ref{SContSemiG} and (d), $A$ is closed
and symmetric. We will show $\ker[A^*\pm \im]=\{0\}$. (This is equivalent
to the self-adjointness of $A$ by the general theorem.) 
Pick $\eta\in \ker[A^*-\im]$.
For all $x\in \h$, one sees
\begin{align}
\frac{\dm}{\dm s}\la T_s x, \eta\ra= -\la AT_s x, \eta\ra=-\la T_s x,
 A^*\eta\ra=
-\im \la T_s x, \eta\ra.
\end{align} 
Here we used the facts $T_s \D(A)\subseteq \D(A)$ and $\displaystyle
\frac{\dm}{\dm s} T_s x =-A T_s x$. Solving the differential equation,
we obtain $\la T_sx, \eta\ra=u_0\,  \ex^{-\im s}$. Since 
\begin{align}
\big|
\la T_s x, \eta\ra
\big|\le \ex^{-s M}\|x\| \|\eta\| \to 0
\end{align} 
as $s\to\infty$, $u_0$ must be $0$. I.e., $\la x, \eta\ra=0$.
Since $\D(A)$ is dense in $\h$, this means $\eta=0$, namely,
$\ker[A^*-\im]=\{0\}$. Similarly we can show $\ker[A^*+\im]=\{0\}$.
Thus we have the assertions in the proposition. $\Box$

\end{document}